\documentstyle[preprint,revtex,eqsecnum]{aps}
\begin{document}
\draft
\begin{title}
Detection, Measurement and Gravitational Radiation
\end{title}
\author{Lee S.~Finn${}^*$}
\begin{instit}
Department of Physics and Astronomy, Northwestern University,
Evanston Illinois\ \ 60208
\end{instit}
\receipt{}
\begin{abstract}

The optimum design, construction, and use of the LIGO, VIRGO, or LAGOS
gravitational radiation detectors depends upon accurate
calculations of their sensitivity to different sources of
radiation.  Here I examine how to determine the sensitivity of
these instruments to sources of gravitational radiation by
considering the process by which data are analyzed in a noisy
detector. The problem of detection (is a signal present in the
output of the detector?) is separated from that of measurement
(what are the parameters that characterize the signal in the
detector output?). By constructing the probability that the
detector output is consistent with the presence of a signal, I
show how to quantify the uncertainty that the output contains a
signal and is not simply noise.  Proceeding further, I construct
the probability distribution that the parameterization
{\boldmath$\mu$} that characterizes the signal has a certain
value. From the distribution and its mode I determine volumes
$V(P)$ in parameter space such that $\mbox{\boldmath$\mu$}\in
V(P)$ with probability $P$ [owing to the random nature of the
detector noise, the volumes $V(P)$ are are always different, even
for identical signals in the detector output], thus quantifying
the uncertainty in the estimation of the signal parameterization.

These techniques are suitable for analyzing the output of a noisy
detector. If we are {\em designing\/} a detector, or determining
the suitability of an existing detector for observing a new
source, then we don't have detector output to analyze but are
interested in the ``most likely'' response of the detector to a
signal. I exploit the techniques just described to determine the
``most likely'' volumes $V(P)$ for detector output that would
result in a parameter probability distribution with given mode.

Finally, as an example, I apply these techniques to determine the
anticipated sensitivity of the LIGO and LAGOS detectors to the
gravitational radiation from a perturbed Kerr black hole.

\end{abstract}

\narrowtext
\section{Introduction}
\label{sec:intro}

Under the present schedule, both the United States Laser Interferometer
Gravitational Wave Observatory (LIGO\cite{ligo91,ligo92}) and the
French/Italian VIRGO \cite{virgo}
will
begin operation in the late 1990s.  Long before that time, theorists must
lay a foundation for the study of gravitational radiation
sources. Part of this foundation involves the construction of
detailed, parameterized models of the waveforms from expected
sources; another part involves the calculation of the anticipated
sensitivity of the detector to each of these sources.
Calculation of these kinds are not only needed for LIGO and VIRGO: design
and technology studies for a Laser Gravitational-wave Observatory
in Space (LAGOS) are currently being pursued\cite{lagos} and
calculations of the sensitivity of LAGOS to appropriate sources
are needed to guide these studies.

In this paper I address the problem of calculating the
anticipated sensitivity of a detector, like LIGO, VIRGO, or LAGOS, to an
arbitrary source of gravitational radiation. The problem breaks
up into two parts which I term {\em detection\/} and {\em
measurement.\/} To ``detect'' is to decide whether the observed
detector output contains a signal from a particular source or is
just an example of noise; to ``measure'' is to assume the
presence of a signal in the detector output and to characterize
the signal in terms of the parameter(s) that describe the source
(and its orientation with respect to the detector).

Echeverria\cite{echeverria} recently examined some of these
issues in the particular context of determining the precision
with which one could characterize the mass and angular momentum
of a perturbed Kerr black hole from observations in a
gravitational radiation detector. The foundation of his analysis
was the construction of a quantity similar to the signal-to-noise
ratio (SNR), and he asserted that the parameters that
characterize a signal observed in the output of the detector are
those that maximized this quantity. This analysis is limited in
two respects (Echeverria \& Finn\cite{ech-finn}):
\begin{enumerate}
\item The validity of the formalism is restricted to the limit of
high SNR; and
\item The formalism cannot determine the amplitude of the signal.
\end{enumerate}
In addition, the conceptual basis of this calculation is not compelling:
the determination of the parameters characterizing a signal in a
noisy detector does not proceed by maximizing the SNR-like
quantity defined by Echeverria\cite{echeverria}.

In contrast, the techniques developed here are all based upon the
construction of probabilities and probability densities. For the
problem of detection, I construct the probability that the
observed detector output is consistent with the presence (or
absence) of a signal. In the case of the measurement problem,
where the detector output is assumed to include a signal, the
quantity constructed is the probability density that describes
the likelihood of a given signal parameterization.

In \S\ref{sec:detection-and-measurement} I examine the twin
processes of detection and measurement from the point of view of
probability theory. The parameters characterizing a signal
identified in the output of a noisy detector are defined to be
those {\em most likely\/} to have resulted in the observed
detector output. Some of the results described in this section
are known elsewhere in the context of data analysis: they are
included here for completeness and so that they may be compared
with the techniques employed in Echeverria\cite{echeverria}.  In
\S\ref{sec:sensitivity} I show how these same techniques can be
exploited to evaluate the {\em anticipated sensitivity\/} of an
instrument to a signal: {\em i.e.,\/} how precisely can the
parameterization of a signal observed in the detector be
determined. I find both exact and, in the interesting limit of a
strong signal, approximate techniques for evaluating the expected
precision with which an observed signal can be described.  As an
example, in \S\ref{sec:bhringdown} I apply the approximate
techniques developed in \S\ref{sec:sensitivity} to the
determination of the parameterization of the gravitational
radiation from a perturbed black hole, especially the black hole
mass $M$ and dimensionless angular momentum parameter $a$. In
\S\ref{sec:discussion} I briefly compare the methods and results
of Echeverria\cite{echeverria} with my own. My conclusions are
presented in \S\ref{sec:conclusions}

\section{Detection, measurement and probability}
\label{sec:detection-and-measurement}

In this section I consider two related problems that arise in the
analysis of the output of a noisy detector: detection and
measurement.  The problem of detection is to determine whether or
not a signal of known form ({\em i.e.,\/} deterministic, though
parameterized by one or several unknown parameters) is present in
the detector output. The problem of measurement is to determine
the values of some or all of the unknown parameters that
characterize the observed signal.

Note that the distinction between detection and measurement
separates the determination of the presence or absence of the
signal from the determination of the parameters that characterize
it: detection does not address the value of the unknown
parameters, and measurement presumes the signal's presence.

Detector noise can always conspire to appear as an example of the
sought-for signal; alternatively, noise can mask the presence of
a signal. In either case, noise interferes with our ability to
determine the presence of the signal or the parameters that
characterize it.  Consequently, any claim of detection must be
associated with a {\em probability\/} signifying the degree of
certainty that the detected signal is not, in fact, an instance
of noise. Similarly, when an observed signal is characterized it
is appropriate to specify both a {\em range\/} of parameters and
a {\em probability\/} that the signal parameters are in the given
range.

For example, I can examine the data stream from a gravitational
radiation detector to determine (with some probability) whether
the radiation from the $l=|m|=2$ mode of a perturbed, rotating
black hole is present, irrespective of the black hole mass,
angular momentum, or orientation with respect to the detector. If
I conclude that a signal is present in the data stream, then I
can attempt to determine bounds on some or all of these
parameters, such that I expect the actual parameters
characterizing the signal to fall within those bounds with a
given probability.

In the next several subsections I examine detection and
measurement in more detail. I assume that the statistical
properties of the detector noise are known, and also that the
{\em form\/} of the sought-for signal is known up to one or
several parameters. My discussion focuses on determining the
probability that a signal of known form is present in the output
of a noisy detector, and on determining the probability that the
unknown parameters have particular values.

While the discussion in \S\ref{sec:bhringdown} is framed in the
context of the measurement of gravitational radiation from
astrophysical sources, the questions addressed in this (and the
following) section are purely statistical ones and contain
nothing that is specific to gravitational radiation, general
relativity, or any particular physical system or theory.  For
more details, the reader may consult Wainstein \&
Zubakov\cite{wainstein}.

\subsection{Detection}
\label{sec:detection}

Consider a data stream $g(t)$ which represents the output of a
detector. The data stream has a noise component $n(t)$ and in
addition may have a signal component $m(t)$. The signal component
is parameterized by several unknown parameters (denoted
collectively as {\boldmath$\mu$}, and individually as $\mu_i$);
hence
\begin{equation}
g(t) = \left\{
\begin{array}{ll}
n(t)&\mbox{if signal not present,}\\
n(t)+m(t;\mbox{\boldmath$\mu$})&\quad
\mbox{if signal $m(t;\mbox{\boldmath$\mu$})$ present.}
\end{array}
\right.\label{defn:g}
\end{equation}
Assume that {\boldmath$\mu$} is continuous, not discrete.
I will describe how to determine the probability that
$m(t;\mbox{\boldmath$\mu$})$,
for {\em undetermined\/} {\boldmath$\mu$}, is
present in $g(t)$, {\em i.e.,\/}
\begin{equation}
P(m|g) \equiv
\left(\begin{tabular}{l}
The conditional probability that a\\
signal of the form $m(t;\mbox{\boldmath$\mu$})$, for\\
unknown \mbox{\boldmath$\mu$}, is present given the\\
observed data stream $g(t)$.
\end{tabular}\right).
\end{equation}

Begin by using Baye's law of conditional probabilities to
re-express $P(m|g)$ as
\begin{equation}
P(m|g) = {P(g|m)P(m)\over P(g)},\label{eqn:pmg}
\end{equation}
where
\begin{mathletters}
\begin{eqnarray}
P(g|m) &\equiv&
\left(\begin{tabular}{l}
The probability of measuring $g$\\
assuming the signal $m$ is present.
\end{tabular}\right),\\
P(m)&\equiv&
\left(\begin{tabular}{l}
The {\em a~priori\/} probability\\
that the signal $m$ is present.
\end{tabular}\right),\\
P(g) &\equiv&
\left(\begin{tabular}{l}
The probability that the\\
data stream $g(t)$ is observed.
\end{tabular}\right).
\end{eqnarray}
\end{mathletters}
Also re-express $P(g)$ in terms of the two possibilities $m$
absent and $m$ present, and further re-express the probability
that $m$ is present in terms of the probability that it is
characterized by the {\em particular\/} {\boldmath$\mu$}:
\begin{eqnarray}
P(g)&=&P(g|0)P(0) + P(g|m)P(m)\nonumber\\
&=&P(g|0)P(0) + P(m)\int d^N\!\mu\,p(\mbox{\boldmath$\mu$})
P[g|m(\mbox{\boldmath$\mu$})],\nonumber\\
&&\label{eqn:unitarity}
\end{eqnarray}
where
\begin{mathletters}
\begin{eqnarray}
P(0) &\equiv&
\left(\begin{tabular}{l}
The {\em a~priori\/} probability that\\
the signal is {\em not\/} present.
\end{tabular}\right),\\
P(g|0) &\equiv&
\left(\begin{tabular}{l}
The probability density of\\
observing $g(t)$ in the absence\\
of the signal.
\end{tabular}\right),\\
P[g|m(\mbox{\boldmath$\mu$})] &\equiv&
\left(\begin{tabular}{l}
The probability density of\\
observing $g(t)$ assuming\\
$m(t;\mbox{\boldmath$\mu$})$ with {\em particular\/}\\
\mbox{\boldmath$\mu$} is present.
\end{tabular}\right),\\
p(\mbox{\boldmath$\mu$}) &\equiv&
\left(\begin{tabular}{l}
The {\em a~priori\/} probability\\
density that $m(t)$ is\\
characterized by \mbox{\boldmath$\mu$}.
\end{tabular}\right).
\end{eqnarray}
\end{mathletters}

Combining equations \ref{eqn:pmg} and \ref{eqn:unitarity}, we
find
\begin{equation}
P(m|g) = {\Lambda\over\Lambda+P(0)/P(m)},\label{eqn:Pmg-Lambda}
\end{equation}
where
\begin{eqnarray}
\Lambda&\equiv&\int d^N\!\mu\,
\Lambda(\mbox{\boldmath$\mu$})\label{defn:lambda}\\
\Lambda(\mbox{\boldmath$\mu$})&\equiv&
p(\mbox{\boldmath$\mu$}){P[g|m(\mbox{\boldmath$\mu$})]\over
P(g|0)}.\label{defn:lambda(t)}
\end{eqnarray}
In equation \ref{eqn:Pmg-Lambda} all of the dependence of
$P(m|g)$ on the data stream $g$ has been gathered into the {\em
likelihood ratio\/} $\Lambda$. Aside from $\Lambda$, $P(m|g)$
depends only on the ratio of the {\em a~priori\/} probabilities
$P(0)$ and $P(m)$. In turn, the likelihood ratio depends on two
components: the {\em a~priori\/} probability density
$p(m|\mbox{\boldmath$\mu$})$ and the ratio
$P[g|m(\mbox{\boldmath$\mu$})]/P(g|0)$.

In order to determine $P(m|g)$ we must {\em assess\/} the {\em
a~priori\/}\ probabilities and {\em calculate\/} the likelihood
ratio.  It is often the case that we know, or can make an
educated guess regarding, the {\em a~priori\/} probabilities. For
example, the sources may be Poisson distributed in time
[determining $P(0)$ and $P(m)$], and they may be homogeneously
distributed in space [determining $p(r)\propto r^2$, where $r$ is
the distance to the source]. At other times our assessment may be
more subjective or based on imperfect knowledge, and in this case
we can use the observed distribution of {\boldmath$\mu$} to test
the validity of our assessments using the techniques of
hypothesis testing (Winkler\cite{winkler} \S7).

Now turn to the evaluation of
$P[g|m(\mbox{\boldmath$\mu$})]/P(g|0)$. To determine this ratio,
first note that the conditional probability of measuring $g(t)$
if the particular signal $m(t;\mbox{\boldmath$\mu$})$ is present
is the same as the conditional probability of measuring $g'(t) =
g(t)-m(t;\mbox{\boldmath$\mu$})$, assuming that the signal
$m(t;\mbox{\boldmath$\mu$})$ is {\em not present\/} in
$g^\prime$:
\begin{equation}
P[g|m(\mbox{\boldmath$\mu$})] = P[g-m(\mbox{\boldmath$\mu$})|0].
\end{equation}
Consequently, we can focus on the conditional probability of
measuring a data stream $g(t)$ under the assumption that no
signal is present [$P(g|0)$].

In the absence of the signal, $g(t)$ is simply an instance of
$n(t)$.  Assume that the $n(t)$ is a normal process with zero
mean, characterized by the correlation function $C_n(\tau)$ [or,
equivalently, by the one-sided power spectral density (PSD)
$S_n(f)$]. In order to compute the ratio
$P[g|m(\mbox{\boldmath$\mu$})]/P(g|0)$, consider the continuum
limit of the case of discretely sampled data $\{g_i: i=1,\ldots,
N\}$, with the correspondence
\begin{mathletters}
\begin{eqnarray}
g_i&=&g(t_i),\\
t_i-t_j&=&(i-j)\Delta t,\\
\Delta t &=& {T\over N-1}.
\end{eqnarray}
\end{mathletters}
The probability that an individual $g_i$ is a sampling of the
random process $n(t)$ is given by
\begin{equation}
P(g_i|0) = {
\exp\left[-{1\over2}{g_i^2\over C_n(0)}\right]\over
\left[2\pi C_n(0)\right]^{1/2}
}
\end{equation}
and the probability that the ordered set $\{g_i: i=1,\ldots N\}$
is a sampling of $n(t)$ is
\begin{equation}
P(g|0) =
{\exp\left[-{1\over2}\sum_{j,k=1}^NC^{-1}_{jk}g_j g_k\right]\over
\left[\left(2\pi\right)^N\det||C_{n,ij}||\right]^{1/2}}
\label{eqn:pg0-discrete}
\end{equation}
where $C^{-1}_{jk}$ is defined by
\begin{equation}
\delta_{jk}\equiv\sum_{l}C_{n,jl}C^{-1}_{lk}
\end{equation}
and
\begin{equation}
C_{n,ij}\equiv C_n\left[\left(i-j\right)\Delta t\right]
\end{equation}
(Mathews \& Walker \cite{mathews} \S14-6, Wainstein \&
Zubakov\cite{wainstein} eqn.~31.11).  Note that the normalization
constant in the denominator of equation \ref{eqn:pg0-discrete} is
independent of the $g_i$; consequently, it does not affect the
{\em ratio\/} $P[g|m(\mbox{\boldmath$\mu$})]/P(g|0)$.  Since it
is this ratio that we are interested in, without loss of
generality drop the normalization constant from the following.

To evaluate equation \ref{eqn:pg0-discrete} in the continuum
limit, first note that
\begin{equation}
\delta(t_j-t_k) =
\lim_{\Delta t\rightarrow0\atop T\rightarrow\infty}
{1\over\Delta t}
\delta_{jk} .
\end{equation}
Consequently,
\widetext
\begin{mathletters}
\begin{eqnarray}
e^{2\pi i f t_k}&=&\sum_j e^{2\pi i ft_j}\delta_{jk}\nonumber\\
&=&\lim_{\Delta t\rightarrow0\atop T\rightarrow\infty}
{1\over\Delta t^2}
\sum_j\Delta t\,e^{2\pi ift_j}
\sum_l\Delta t\,C_n(t_j-t_l)C^{-1}(t_l,t_k)\nonumber\\
&=&\lim_{\Delta t\rightarrow0\atop T\rightarrow\infty}
{1\over\Delta t^2}
\int_{-\infty}^{\infty} dt_j\,e^{2\pi ift_j}
\int_{-\infty}^{\infty} dt_l\,C_n(t_j-t_l)C^{-1}(t_l,t_k)\nonumber\\
&=&\lim_{\Delta t\rightarrow0}{1\over\Delta t^2}
\int_{-\infty}^{\infty} dt_l\,e^{2\pi ift_l} C^{-1}(t_l,t_k)
\int_{-\infty}^{\infty} d\tau\,e^{2\pi if\tau}C_n(\tau)
\label{eqn:use-wk}\\
&=&\lim_{\Delta t\rightarrow0}{1\over\Delta t^2}
{1\over 2}S_n(f){\widetilde{C^{-1}}}(f,t_k).
\label{eqn:wk-used}
\end{eqnarray}
\end{mathletters}
\narrowtext
To proceed from equation \ref{eqn:use-wk} to \ref{eqn:wk-used},
use the Wiener-Khintchine ({\em cf.\/} Kittel\cite{kittel} \S28)
theorem to relate the PSD $S_n(f)$ to the correlation function
$C_n(\tau)$ and define
\begin{equation}
{\widetilde{C^{-1}}}(f,t_k) \equiv
\int_{-\infty}^\infty dt\,C^{-1}(t,t_k)e^{2\pi ift}.
\end{equation}
Consequently, as we approach the continuum limit, we have
\begin{equation}
{\widetilde{C^{-1}}}(f,t_k) =  \lim_{\Delta t\rightarrow0}\Delta t^2
{2e^{2\pi ift_k}\over S_n(f)}.
\end{equation}
With $\widetilde{C^{-1}}$ and Parseval's Theorem, we can evaluate
the continuum limit of the argument of the exponential in
equation
\ref{eqn:pg0-discrete}:
\widetext
\begin{eqnarray}
\lim_{\Delta t\rightarrow0\atop T\rightarrow\infty}
\sum_{j,k=1}^N C^{-1}_{jk}g_jg_k&=&
\lim_{\Delta t\rightarrow0\atop T\rightarrow\infty}
{1\over\Delta t^2}\sum_{j,k=1}^N
\Delta t^2 C^{-1}(t_j,t_k)g(t_j)g(t_k)\nonumber\\
&=&\lim_{\Delta t\rightarrow0}{1\over\Delta t^2}
\int\!\!
\int_{-\infty}^\infty dt_j\,dt_k\,
C^{-1}(t_j,t_k)g(t_j)g(t_k)\nonumber\\
&=&\lim_{\Delta t\rightarrow0}{1\over\Delta t^2}
\int\!\!
\int_{-\infty}^\infty
df\,dt_k\,
{\widetilde{C^{-1}}}(f,t_k){\widetilde{g}}^*(f)g(t_k)\nonumber\\
&=&2\int_{-\infty}^\infty df\,
{{\widetilde{g}}^*(f)\over S_n(|f|)}
\int_{-\infty}^\infty dt_k\,e^{2\pi ift_k}g(t_k)
\nonumber\\
&=&2\int_{-\infty}^\infty df\,
{{\widetilde{g}}(f){\widetilde{g}}^*(f)\over S_n(|f|)}.
\end{eqnarray}
\narrowtext
Here and henceforth we will denote the Fourier transform of
$r(t)$ as $\widetilde{r}(f)$.

Since the detector output $g(t)$ is real,
${\widetilde{g}}^*(f)={\widetilde{g}}(-f)$. Define the
symmetric inner product $\left<g,h\right>$
\begin{equation}
\left<g,h\right>\equiv\int_{-\infty}^\infty df\,
{{\widetilde{g}(f)}{\widetilde{h}^*(f)}\over S_n(|f|)}.
\label{eqn:inner-prod}
\end{equation}
for real functions $g$ and $h$.  In terms of this inner product,
\begin{eqnarray}
\Lambda(\mbox{\boldmath$\mu$}) &=&
p(\mbox{\boldmath$\mu$}){P[g|m(\mbox{\boldmath$\mu$})]
\over P(g|0)}\nonumber\\
&=& p(\mbox{\boldmath$\mu$})\exp\left[
2\left<g,m(\mbox{\boldmath$\mu$})\right> -
\left<m(\mbox{\boldmath$\mu$}),m(\mbox{\boldmath$\mu$})\right>
\right],\label{eqn:likelihood-ratio}
\end{eqnarray}
The likelihood ratio $\Lambda$ is found by substituting eqn.\
\ref{eqn:likelihood-ratio} into eqn.\ \ref{defn:lambda}.

To summarize, the probability $P(m|g)$ that a signal of the class
$m(t;\mbox{\boldmath$\mu$})$ is present in the output of the
detector $g(t)$ can be expressed in terms of three {\em
a~priori\/} probabilities [$P(0)$, $P(m)$, and
$p(\mbox{\boldmath$\mu$})$] and the ratio of two conditional
probabilities [$P(m|g)/P(0|g)$]. The {\em a~priori\/}
probabilities must be assessed, while the ratio of the
conditional probabilities can be calculated. Often we know, or
can make an educated guess regarding, the {\em a~priori\/}
probabilities; at other times our assessment is subjective or
otherwise based on imperfect knowledge.  Finally we establish a
threshold for $P(m|g)$ [or, equivalently, for $\Lambda$,
$\ln\Lambda$, or some other surrogate of $P(m|g)$], and say that
if the $P(m|g)$ (or its surrogate) exceeds the threshold then we
have detected the signal.

I will not discuss detection further, except to say that the
choice of threshold is influenced by our strategy to minimize
errors. The two kinds of errors we can make are to claim the
presence of a signal when one is in fact not present (a ``false
alarm''), or to dismiss an observed $g(t)$ as noise when a signal
is present (a ``false dismissal''). In order to minimize the
probability of a false alarm (conventionally denoted $\alpha$) we
want a large threshold, while to minimize the probability of a
false dismissal (conventionally denoted $\beta$) we want a small
threshold. One obvious strategy for choosing the threshold is to
minimize the sum $\alpha+\beta$, {\em i.e.,\/} to minimize the
probability of making an error. Alternatively, some other
combination of $\alpha$ and $\beta$ may be minimized, taking
into account the relative seriousness of the different kinds of
errors.  Regardless, it is inadvisable to blindly choose a
threshold for $P(m|g)$ without careful consideration of the false
alarm and false dismissal probabilities that arise and their
relative severity.

\subsection{Measurement}
\label{sec:measurement}

Turn now to the question of measurement. From equations
\ref{eqn:pmg}, \ref{eqn:unitarity} and \ref{defn:lambda(t)} we
have
\begin{eqnarray}
p[m(\mbox{\boldmath$\mu$})|g] &=&
\left(\begin{tabular}{l}
The conditional probability that\\
the particular signal $m(t;\mbox{\boldmath$\mu$})$ is\\
present in the data stream $g(t)$.
\end{tabular}\right)\nonumber\\
&=&
{\Lambda(\mbox{\boldmath$\mu$})\over\Lambda+P(0)/P(m)}
\label{eqn:p[m(mu)|g]}.
\end{eqnarray}
This conditional probability density is directly proportional to
$\Lambda(\mbox{\boldmath$\mu$})$ and, since the denominator in
equation \ref{eqn:p[m(mu)|g]} is independent of {\boldmath$\mu$},
it is maximized where $\Lambda(\mbox{\boldmath$\mu$})$ is
maximized. If we assume that the signal is present, then the
probability density that it is characterized by
\mbox{\boldmath$\mu$} is
\begin{equation}
p[m(\mbox{\boldmath$\mu$})|g,m] =
{\Lambda(\mbox{\boldmath$\mu$})\over\Lambda}.
\label{eqn:p[m(mu)|g]-assuming}
\end{equation}

The goal of the measurement process is to determine a volume
$V(P)$ in parameter space such that $\mbox{\boldmath$\mu$}\in
V(P)$ with probability $P$.  This volume is ``centered'' on the
mode of the distribution $p[m(\mbox{\boldmath$\mu$})|g]$ in a way
we define later on. The mode of either
$p[m(\mbox{\boldmath$\mu$})|g]$ or
$p[m(\mbox{\boldmath$\mu$})|g,m]$ is the {\boldmath$\mu$} that
maximizes $\Lambda(\mbox{\boldmath$\mu$})$.  Denote the mode by
$\widehat{\mbox{\boldmath$\mu$}}$.\footnote{While we assume in
what follows that the distribution has a single mode, the
generalization to a multi-modal distribution is trivial.}  While
I will occasionally refer to $\widehat{\mbox{\boldmath$\mu$}}$ as
the ``measured'' parameterization of the signal, bear in mind
that $\widehat{\mbox{\boldmath$\mu$}}$ is only the {\em most
likely\/} parameterization of the observed signal.

If we assume that the global maximum of
$\Lambda(\mbox{\boldmath$\mu$})$ is also a local extremum, then
$\widehat{\mbox{\boldmath$\mu$}}$ satisfies
\begin{equation}
0 = {\partial\Lambda(\mbox{\boldmath$\mu$})\over\partial\mu_i};
\label{eqn:exp-non-linear-extrema}
\end{equation}
equivalently,
$\widehat{\mbox{\boldmath$\mu$}}$ maximizes
\begin{equation}
\ln\Lambda(\mbox{\boldmath$\mu$}) =
\ln p(\mbox{\boldmath$\mu$})+
2\left<m(\mbox{\boldmath$\mu$}),g\right> -
\left<m(\mbox{\boldmath$\mu$}),m(\mbox{\boldmath$\mu$})\right>,
\end{equation}
{\em i.e.,\/} it satisfies
\begin{equation}
0 = {\partial\ln p(\widehat{\mbox{\boldmath$\mu$}})\over\partial\mu_i}+
2\left<{\partial
m\over\partial\mu_i}(\widehat{\mbox{\boldmath$\mu$}}),
g-m(\widehat{\mbox{\boldmath$\mu$}})\right> .
\label{eqn:non-linear-extrema}
\end{equation}
This final set of equations is in general non-linear and may be
satisfied by several different {\boldmath$\mu$}.  Some will
represent local maxima, while others will correspond to local
minima or inflection points; thus,
equation~\ref{eqn:exp-non-linear-extrema} is a necessary but not
sufficient condition for $\widehat{\mbox{\boldmath$\mu$}}$.

An important characterization of the strength of the signal in a
detector is the signal-to-noise ratio (SNR).  The ``actual'' SNR
depends on the true parameterization of the signal
$\widetilde{\mbox{\boldmath$\mu$}}$.  We do not have access to
$\widetilde{\mbox{\boldmath$\mu$}}$; however, we {\em do\/} know
that the most likely value of $\widetilde{\mbox{\boldmath$\mu$}}$
is $\widehat{\mbox{\boldmath$\mu$}}$, and we define the SNR in
terms of $\widehat{\mbox{\boldmath$\mu$}}$:
\begin{equation}
\rho^2 = 2\left<m(\widehat{\mbox{\boldmath$\mu$}}),
m(\widehat{\mbox{\boldmath$\mu$}})\right>.\label{defn:snr}
\end{equation}
The factor of two arises because the power spectral density
$S_n(f)$ is one-sided while $\widetilde{m}(f)$ is two-sided. Note
that $\rho^2$ is expressed in terms of the signal power ({\em
i.e.,\/} it is proportional to the square of the signal
amplitude). There is some ambiguity in the literature over
whether ``SNR'' refers to $\rho$ or $\rho^2$. We avoid the
ambiguity by referring to either $\rho$ or $\rho^2$ wherever the
context demands it.

Having found the distribution $p[m(\mbox{\boldmath$\mu$})|g]$ (or
$p[m(\mbox{\boldmath$\mu$})|g,m]$), we
define the boundary of the volumes $V(P)$ to be its iso-surfaces.
The probability $P$
corresponding to the iso-surface
$p[m(\mbox{\boldmath$\mu$})|g,m]=K^2$ is
\begin{equation}
P = \int_{p[m(\mbox{\boldmath$\scriptsize\mu$})|g,m]\geq K^2} d^N\!\mu\,
p[m(\mbox{\boldmath$\mu$})|g].\label{eqn:P-from-K}
\end{equation}

Note that since the distribution
$p[m(\mbox{\boldmath$\mu$})|g,m]$ is not generally symmetric,
$\widehat{\mbox{\boldmath$\mu$}}$ is not necessarily the {\em
mean\/} of {\boldmath$\mu$}. Also, if the distribution
$p[m(\mbox{\boldmath$\mu$})|g]$ has more than one local maximum
then $V(P)$ need not be simply connected.

To summarize, suppose we have an observation $g(t)$ which we
assume (or conclude) includes a signal
$m(\widetilde{\mbox{\boldmath$\mu$}})$ (for unknown
$\widetilde{\mbox{\boldmath$\mu$}}$). We construct the
probability density $p[m(\mbox{\boldmath$\mu$})|g,m]$ according
to equation \ref{eqn:p[m(mu)|g]-assuming}, and identify
iso-surfaces of $p[m(\mbox{\boldmath$\mu$})|g,m]$ as the boundary
of probability volumes $V(P)$ according to equation
\ref{eqn:P-from-K}.  Finally, we assert that
$\widetilde{\mbox{\boldmath$\mu$}}\in V(P)$ with probability $P$.

\section{Measurement sensitivity}
\label{sec:sensitivity}

In \S\ref{sec:detection-and-measurement} we saw how to decide
whether a signal is present or absent from the output of a noisy
detector, and, if present, how to determine bounds on the
parameterization of the signal. Now I show how to {\em
anticipate\/} the precision with which a detector can place
bounds on the parameterization that characterizes a signal. In
particular, consider an observed $g(t)$ which contains a signal
$m(t; \widetilde{\mbox{\boldmath$\mu$}})$ for unknown
$\widetilde{\mbox{\boldmath$\mu$}}$. We are interested ultimately
in the distribution of
\begin{equation}
\delta\mbox{\boldmath$\mu$} \equiv
\widetilde{\mbox{\boldmath$\mu$}}-\widehat{\mbox{\boldmath$\mu$}}
\label{defn:delta-mu}
\end{equation}
where $\widehat{\mbox{\boldmath$\mu$}}$ is determined by the
techniques discussed in \S\ref{sec:detection-and-measurement}.
There are an infinity of possible $g(t)$ that can lead to the
same $\widehat{\mbox{\boldmath$\mu$}}$ [corresponding to
different instances of the noise $n(t)$], and for each there is a
different probability distribution
$p[m(\mbox{\boldmath$\mu$})|g]$ ({\em cf.\/}
eqn.~\ref{eqn:p[m(mu)|g]}) and a different set of probability
volumes $V(P)$. We will find the probability volumes $V(P)$
corresponding to
\begin{equation}
p(\widetilde{\mbox{\boldmath$\mu$}}|
\widehat{\mbox{\boldmath$\mu$}}) =
\left(\begin{tabular}{l}
The conditional probability density\\
that the signal parameterization is\\
$\widetilde{\mbox{\boldmath$\mu$}}$,
assuming that the mode of the\\
distribution
$p[m(\mbox{\boldmath$\mu$})|g]$ is
$\widehat{\mbox{\boldmath$\mu$}}$.
\end{tabular}\right).
\end{equation}
I show first how to do this exactly, and then show a useful
approximation for strong signals.

The mode $\widehat{\mbox{\boldmath$\mu$}}$ of the distribution
$p[m(\mbox{\boldmath$\mu$})|g,m]$ satisfies
\begin{eqnarray}
2\left<
m(\widetilde{\mbox{\boldmath$\mu$}}) -
m(\widehat{\mbox{\boldmath$\mu$}}),
{\partial m\over\partial\mu_j}
(\widehat{\mbox{\boldmath$\mu$}})
\right>\nonumber\\
\qquad{}+{\partial\ln p\over\partial\mu_j}
(\widehat{\mbox{\boldmath$\mu$}})
&=&
-2\left<
n,{\partial m\over\partial\mu_j}
(\widehat{\mbox{\boldmath$\mu$}})
\right>
\label{eqn:linearize-me}
\end{eqnarray}
({\em cf.\/} eqns.~\ref{defn:g} and
\ref{eqn:non-linear-extrema}).  Since $n(t)$ is a normal variable
with zero mean, so are each of the $\left<n,\partial
m/\partial\mu_j\right>$ on the righthand side of
equation~\ref{eqn:linearize-me}. Denote these random variables
$\nu_i$:
\begin{equation}
\nu_i \equiv
2\left<n,
{\partial m\over\partial\mu_i}(\widehat{\mbox{\boldmath$\mu$}})
\right> .
\end{equation}
The joint distribution of the $\nu_i$ is a multivariate Gaussian
and its properties determine, through the
equation~\ref{eqn:linearize-me}, the properties of the
distribution of $\delta\mbox{\boldmath$\mu$}$.  Consequently we
can focus on the joint distribution of the $\nu_i$.

Since the $\nu_i$ are normal, their distribution is determined
completely by the means $\overline{\nu_i}$, which vanish, and the
quadratic moments
\begin{equation}
\overline{\nu_i\nu_j} =
4\overline{
\left<n,{\partial m\over\partial\mu_i}
(\widehat{\mbox{\boldmath$\mu$}})\right>
\left<n,{\partial m\over\partial\mu_j}
(\widehat{\mbox{\boldmath$\mu$}})\right>
}.
\end{equation}
To evaluate the average on the righthand side, we will use the
ergodic principle to turn the ensemble average over the random
process $n$ into a time average over a particular instance of
$n$. Recalling that a time translation affects the Fourier
transform of a function by a change in phase,
\begin{equation}
{\cal F}\left[r(t+\tau)\right] = e^{-2\pi
if\tau}{\cal F}\left[r(t)\right],
\end{equation}
write
\begin{equation}
\left<n(t+\tau),r(t)\right> =
\int_{-\infty}^\infty dt\,
e^{-2\pi if\tau}{\widetilde{n}(f)\widetilde{r}^*(f)\over S_n(f)}.
\end{equation}
Consequently,
\widetext
\begin{mathletters}
\begin{eqnarray}
\overline{\left<n,r\right>\left<n,s\right>} &=&
\lim_{T\rightarrow\infty}{1\over2T}
\int_{-T}^{T}d\tau\,
\left<n(t+\tau),r\right>\left<n(t+\tau),s\right>\nonumber\\
&=&
\lim_{T\rightarrow\infty}{1\over2T}
\int_{-T}^{T}d\tau\,
\int_{-\infty}^\infty df\,
  {\widetilde{n}(f)\widetilde{r}^*(f)\over S_n(f)}
  e^{-2\pi if\tau}
  \int_{-\infty}^\infty df'\,
  {\widetilde{n}(f')\widetilde{s}^*(f')\over S_n(f')}
  e^{-2\pi if'\tau}\nonumber\\
&=&
\lim_{T\rightarrow\infty}{1\over2T}
\int_{-\infty}^\infty df\,
  {\widetilde{n}(f)\widetilde{r}^*(f)\over S_n(f)}
\int_{-\infty}^\infty df'\,
  {\widetilde{n}(f')\widetilde{s}^*(f')\over S_n(f')}
  \delta(f+f')\nonumber\\
&=&
\lim_{T\rightarrow\infty}{1\over2T}
\int_{-\infty}^\infty df\,
  {\widetilde{n}(f)\widetilde{n}^*(f)\over S_n(f)}
  {\widetilde{r}^*(f)\widetilde{s}^*(f)\over S_n(f)}
\label{eqn:before-psd}\\
&=&
{1\over 2}\int_{-\infty}^\infty df\,
  {\widetilde{r}^*(f)\widetilde{s}^*(f)\over S_n(f)}
\label{eqn:after-psd}\\
&=&{1\over2}\left<r,s\right>\label{eqn:quad-delta}.
\end{eqnarray}
\end{mathletters}
\narrowtext
In going from eqn.\ \ref{eqn:before-psd} to eqn.\
\ref{eqn:after-psd}, we used the definition of the PSD of the
detector noise $n(t)$:
\begin{equation}
S_n(f) \equiv
\lim_{T\rightarrow\infty}{1\over T} \left|\widetilde{n}(f)\right|^2
\end{equation}
({\em cf.\/} Kittel\cite{kittel} \S28). With the result in
equation
\ref{eqn:quad-delta}, we have
\begin{mathletters}
\begin{eqnarray}
\overline{\nu_i\nu_j} &=& 4\overline{
\left<n,{\partial m\over\partial\mu_i}
(\widehat{\mbox{\boldmath$\mu$}})\right>
\left<n,{\partial m\over\partial\mu_j}
(\widehat{\mbox{\boldmath$\mu$}})\right>
}\nonumber\\
&=&
2\left<
{\partial m\over\partial\mu_i}
(\widehat{\mbox{\boldmath$\mu$}}),
{\partial m\over\partial\mu_j}
(\widehat{\mbox{\boldmath$\mu$}})
\right>\\
&\equiv&{\cal C}^{-1}_{ij}.\label{defn:cij}
\end{eqnarray}
\end{mathletters}
In terms of the ${\cal C}_{ij}$ ({\em i.e.,\/} the inverse of the
${\cal C}^{-1}_{ij}$), the joint distribution of the $\nu_i$ is
given by
\begin{equation}
p(\mbox{\boldmath$\nu$}) =
{\exp\left[-{1\over2}\sum_{i,j}{\cal C}_{ij}\nu_i\nu_j\right]\over
\left[\left(2\pi\right)^N\det||{\cal C}^{-1}_{ij}||\right]^{1/2}
}\label{eqn:mv-gaussian}
\end{equation}
This is also the joint distribution of the quantities
\begin{equation}
- 2\left<
m(\widetilde{\mbox{\boldmath$\mu$}}) -
m(\widehat{\mbox{\boldmath$\mu$}}),
{\partial m\over\partial\mu_j}(\widehat{\mbox{\boldmath$\mu$}})
\right> -
{\partial\ln p\over\partial\mu_j}
(\widehat{\mbox{\boldmath$\mu$}})
\end{equation}
that appear on the lefthand side of
equation~\ref{eqn:linearize-me}; consequently, we expect that for
an observation characterized by a given
$\widehat{\mbox{\boldmath$\mu$}}$ the probability volumes $V(P)$
are given implicitly by
\FL
\begin{eqnarray}
K^2&\geq&
\sum_{i,j}{\cal C}_{ij}\left[
2\left<
m(\widetilde{\mbox{\boldmath$\mu$}}) -
m(\widehat{\mbox{\boldmath$\mu$}}),
{\partial m\over\partial\mu_i}(\widehat{\mbox{\boldmath$\mu$}})
\right> +
{\partial\ln p\over\partial\mu_i}
(\widehat{\mbox{\boldmath$\mu$}})
\right]\nonumber\\
&&\quad{}\times
\left[
2\left<
m(\widetilde{\mbox{\boldmath$\mu$}})
- m(\widehat{\mbox{\boldmath$\mu$}}),
{\partial m\over\partial\mu_j}(\widehat{\mbox{\boldmath$\mu$}})
\right> +
{\partial\ln p\over\partial\mu_j}
(\widehat{\mbox{\boldmath$\mu$}})
\right]\nonumber\\
&&\label{eqn:exact-dist}
\end{eqnarray}
where
\begin{equation}
P = \int_{\sum_{i,j}{\cal C}_{ij}\nu_i\nu_j\leq K^2} d^N\nu
{\exp\left[-{1\over2}\sum_{i,j}{\cal C}_{ij}\nu_i\nu_j\right]\over
\left[\left(2\pi\right)^N\det||{\cal C}^{-1}_{ij}||\right]^{1/2}
}.\label{eqn:exact-dist-P}
\end{equation}
This result is exact as long as the maximum
$\widehat{\mbox{\boldmath$\mu$}}$ of
$\Lambda(\mbox{\boldmath$\mu$})$ is also a local extremum of
$\Lambda(\mbox{\boldmath$\mu$})$.

As the SNR becomes large the distribution
$p(\widetilde{\mbox{\boldmath$\mu$}}|
\widehat{\mbox{\boldmath$\mu$}})$
becomes sharply peaked about $\widehat{\mbox{\boldmath$\mu$}}$
and the determination of the volume $V(P)$ is greatly simplified.
Suppose that $\rho^2$ is so large that for
$\widetilde{\mbox{\boldmath$\mu$}}\in V(P)$ for all $P$ of
interest, the difference $m(\widetilde{\mbox{\boldmath$\mu$}})-
m(\widehat{\mbox{\boldmath$\mu$}})$ can be linearized in
$\delta\mbox{\boldmath$\mu$}$. We then obtain in place of
equation~\ref{eqn:linearize-me}
\begin{equation}
\sum_{i}\delta\mu_i\,{\cal C}_{ij}^{-1}
= -2\left<n,
{\partial m\over\partial\mu_j}(\widehat{\mbox{\boldmath$\mu$}})
\right> -
{\partial\ln p\over\partial\mu_j}(\widehat{\mbox{\boldmath$\mu$}})
\label{eqn:delta-mu-j}
\end{equation}
The random variables $\delta\mbox{\boldmath$\mu$}$ are related to
the {\boldmath$\nu$} by a linear transformation,
\begin{equation}
\delta\mu_i =
-\sum_{j}{\cal C}_{ij}\left[\nu_j+
{\partial\ln p\over\partial\mu_j}(\widehat{\mbox{\boldmath$\mu$}})
\right];
\end{equation}
consequently, the $\delta\mbox{\boldmath$\mu$}$ are normal with
means
\begin{equation}
\overline{\delta\mu_i} =
-\sum_{j}{\cal C}_{ij}{\partial\ln p\over\partial\mu_j}
(\widehat{\mbox{\boldmath$\mu$}}),
\end{equation}
and quadratic moments
\begin{eqnarray}
\overline{
\left(\delta\mu_i-\overline{\delta\mu_i}\right)
\left(\delta\mu_j-\overline{\delta\mu_j}\right)}
= {\cal C}_{ij}.
\end{eqnarray}
The probability distribution
$p(\delta\mbox{\boldmath$\mu$}|\widehat{\mbox{\boldmath$\mu$}})$
is a multivariate Gaussian ({\em cf.\/}
eqn.~\ref{eqn:mv-gaussian}):
\FL
\begin{equation}
p(\delta\mbox{\boldmath$\mu$}|\widehat{\mbox{\boldmath$\mu$}}) =
{\exp\left[
-{1\over2}\sum_{i,j}{\cal C}^{-1}_{ij}
\left(\delta\mu_i-\overline{\delta\mu_i}\right)
\left(\delta\mu_j-\overline{\delta\mu_j}\right)
\right]\over
\left[\left(2\pi\right)^N\det||{\cal C}_{ij}||\right]^{1/2}}.
\label{eqn:prob-dist}
\end{equation}
Note that the matrix ${\cal C}_{ij}$ now has acquired a physical
meaning: in particular, we see that the variances $\sigma^2_i$ of
the $\delta\mu_i$ are
\begin{eqnarray}
\sigma^2_i &\equiv& \overline{
\left(\delta\mu_i-\overline{\delta\mu_i}\right)^2}\nonumber\\
&=& {\cal C}_{ii}\label{defn:sigma}
\end{eqnarray}
and the correlation coefficients $r_{ij}$ are given by
\begin{eqnarray}
r_{ij} &\equiv&\sigma_i^{-1}\sigma_j^{-1}
\overline{
\left(\delta\mu_i-\overline{\delta\mu_i}\right)
\left(\delta\mu_j-\overline{\delta\mu_j}\right)}
\nonumber\\
&=&{{\cal C}_{ij}\over\sigma_i\sigma_j}
\label{defn:correlation}
\end{eqnarray}
In this sense we say that ${\cal C}_{ij}$ is the covariance
matrix of the random variables $\delta\mbox{\boldmath$\mu$}$.

In the strong signal approximation, the surfaces bounding the
volume $V(P)$ are ellipsoids defined by the equation
\begin{equation}
\sum_{i,j}\left(\delta\mu_i-\overline{\delta\mu_i}\right)
\left(\delta\mu_j-\overline{\delta\mu_j}\right)
{\cal C}^{-1}_{ij} = K^2,
\end{equation}
where the constant $K^2$ is related to $P$ by
\begin{equation}
P=\int_{\sum_{i,j}{\cal C}^{-1}_{ij}x^ix^j\leq K^2}
d^N\! x\,
{\exp\left[-{1\over2}\sum_{i,j}{\cal C}^{-1}_{ij}x^ix^j\right]
\over\left[\left(2\pi\right)^N\det||{\cal C}_{ij}||\right]^{1/2}}
\end{equation}

It is often the case that not all of the parameters that
characterize the signal are of physical interest. In that case,
we may integrate the probability distribution
(eqn.~\ref{eqn:mv-gaussian} or \ref{eqn:prob-dist}) over the
uninteresting parameters, leaving a distribution describing just
the parameters of physical interest.

Finally we come to the question of when the linearization in
equation \ref{eqn:delta-mu-j} is a reasonable approximation. Two
considerations enter here:
\begin{enumerate}
\item It is important that the probability contours of interest
({\em e.g.,\/} 90\%) do not involve $\delta\mbox{\boldmath$\mu$}$
so large that the linearization of
$m(\widetilde{\mbox{\boldmath$\mu$}}) -
m(\widehat{\mbox{\boldmath$\mu$}})$ is a poor approximation; and
\item It is important that the condition number ({\em cf.\/}
Golub \& Van Loan\cite{golub}) of the matrix ${\cal C}^{-1}_{ij}$
be sufficiently small that the inverse ${\cal C}_{ij}$ is
insensitive to this approximation in the neighborhood of
$\widehat{\mbox{\boldmath$\mu$}}$.\footnote{Recall that the
relative error in $\delta\mbox{\boldmath$\mu$}$ is the condition
number times the relative error in ${\cal C}_{ij}^{-1}$: for a
large condition number, small errors in ${\cal C}_{ij}^{-1}$
introduced by the linearization approximation can result in large
errors in $\delta\mbox{\boldmath$\mu$}$.}
\end{enumerate}
These two conditions will depend on the problem addressed.  If
the validity of the linearization procedure is doubtful owing to
the violation of either or both of these conditions, then we must
fall-back on equation~\ref{eqn:linearize-me} and the exact
results in equations~\ref{eqn:exact-dist} and
\ref{eqn:exact-dist-P}.

\section{Application: A perturbed black hole}
\label{sec:bhringdown}

In this section, I show how to use the approximate techniques
developed in \S\ref{sec:sensitivity} to find the precision with
which the mass and angular momentum of a perturbed black hole can
be determined through measurement in an interferometric
gravitational wave detector. This problem was first considered by
Echeverria\cite{echeverria}.

Consider a single interferometric gravitational wave detector and
a perturbed black hole of mass $M$ and dimensionless angular
momentum parameter $a$.  Focus attention on a single oscillation
mode of the black hole, {\em e.g.,\/} the $l=m=2$ mode. The
strain measured by the detector has the time dependence of an
exponentially damped sinusoid characterized by the four
parameters $Q$, $f$, $V$, and $T$:
\FL
\begin{equation}
h(t) = \left\{\begin{array}{ll}
0&\quad\mbox{for $t<0$},\\
V^{-1/3}
e^{-\pi f(t-T)/Q}
\sin\left[2\pi f(t-T)\right]&
\quad\mbox{for $t>0$}.
\end{array}\right.
\label{defn:h}
\end{equation}
For convenience, assume that the perturbation begins abruptly at
the {\em starting time\/} $T$. The {\em frequency\/} $f$ depends
inversely on the mass of the black hole, and has a weak
dependence on its angular momentum: for the $l=m=2$ quasi-normal
oscillation mode,
\begin{mathletters}
\begin{eqnarray}
f &\simeq& {F(a)\over2\pi M}\label{eq:f}\\
F(a) &\equiv& 1-{63\over100}(1-a)^{3/10},\label{eqn:f(a)}
\end{eqnarray}
\end{mathletters}
is an accurate semi-empirical expression for the real part of the
quasi-normal mode frequency (Echeverria\cite{echeverria} eqn.~4.4
and tbl.~II). The {\em quality\/} $Q$ is the damping time $\tau$
measured in units of the frequency $f$:
\begin{equation}
Q = \pi f\tau.\label{defn:q}
\end{equation}
For the $l=m=2$ oscillation mode of the black hole, $Q$ depends
entirely on $a$:
\begin{equation}
Q \simeq Q(a) \equiv 2(1-a)^{-9/20} \label{eqn:q(a)}
\end{equation}
(Echeverria\cite{echeverria} eqn.~4.3 and tbl.~II). Finally, the
{\em amplitude\/} $V^{-1/3}$ of the waveform depends on the
distance to the source, the size of the perturbation, and the
relative orientation of the detector and the source.

This peculiar parameterization of the amplitude reflects our
expectation that perturbed black holes are distributed uniformly
throughout space ({\em i.e.,\/} $V\propto r^3$) and that all
relative orientations of the detector and the black hole source
are equally probable.  Additionally, it reflects an assumption
that perturbations of any allowed amplitude are equally probable;
consequently, the {\em a~priori\/} distribution $p(V)$ is
uniform.  Let us also assume that $p(a)$, $p(f)$ and $p(T)$ are
uniform and that there is no {\em a~priori\/} correlation of $a$,
$f$, $V$, or $T$.

An interferometric gravitational wave detector is naturally a
broadband receiver, though it can be operated in a narrow band
mode ({\em cf.\/} Vinet, Meers, Man, \& Brillet \cite{vinet},
Meers\cite{meers} and Krolak, Lobo \& Meers\cite{klm}). Assume
that the detector response function is uniform in the frequency
domain over the bandwidth of the gravitational wave;
consequently, the signal component in the output of the detector
[$m(t,\mbox{\boldmath$\mu$})$] is equal to the waveform
$h(t;Q,f,V,T)$ ({\em cf.\/} eqn.~\ref{defn:h}).  Assume also that
the noise PSD ($S_n$) of the detector is independent of frequency
($f$) in the bandwidth ($1/\tau$) of the signal (I will discuss
the validity of this approximation below).

\subsection{The signal-to-noise ratio}

As a first step toward finding the precision with which $a$, $M$,
$V$ and $T$ can be determined, we calculate the SNR $\rho^2$.
With $h$ given by equation~\ref{defn:h}, evaluate $\rho^2$ using
equation~\ref{defn:snr} to obtain
\begin{equation}
\rho^2 = {2 Q^3\over\pi f V^{2/3}(1+4Q^2) S_n}.\label{eqn:snr}
\end{equation}
This expression is valid to better than a percent as long as the
signal is observed for a period of time $\Delta t\agt2.5\tau$.

In arriving at equation \ref{eqn:snr} we assumed that the noise
PSD is constant over the bandwidth of the signal so that
$S_n=S_h(f)$. The signal bandwidth $\Delta f$ is approximately
\begin{equation}
\Delta f \simeq {1\over\tau} = {\pi\over2}\left(1-a\right)^{9/20} f.
\end{equation}
For small $a$ the bandwidth is approximately $f$, while for large
$a$ the signal is monochromatic. For small $a$ the approximation
that $S_n$ is constant over the bandwidth of the signal is only a
fair approximation for LIGO ({\em cf.\/} Vogt\cite{ligo91},
Abramovici {\em et al.\/}\cite{ligo92}) or LAGOS ({\em cf.\/}
Faller {\em et al.\/}\cite{lagos}); however, it becomes a good
approximation for both detectors when $a\agt0.9$ (corresponding
to $\Delta f/f\alt1/2$).

The amplitude $V$ depends on the detector orientation with
respect to the black hole, the amplitude of the perturbation, and
the distance between the black hole and the detector. Average
$\rho^2$ over all possible orientations of the detector with
respect to the black hole ({\em cf.\/} Thorne\cite{300yrs}
\S9.5.3) to obtain
\begin{equation}
\overline{\rho^2}=
{16\over5}
{Q^2\over F^2\left(1+4Q^2\right)}
{\epsilon M\over S_n}
\left(M\over r\right)^2\label{eqn:rho-epsilon}
\end{equation}
where $\epsilon M$ is the total energy radiated by the $l=m=2$
mode of the black hole perturbation and $r$ is the distance of
the source.

\widetext
When operated as a broadband detector, the LIGO advanced
detectors will be most sensitive to perturbed black holes with
$50\,{\rm M}_\odot\alt M\alt100\,{\rm M}_\odot$ where
$S_n\simeq10^{-48}\,{\rm Hz}^{-1}$ ({\em cf.\/} Krolak, Lobo \&
Meers\cite{klm}, Dhurandhar, Krolak \& Lobo\cite{dhurandhar},
Vogt {\em et al.}\cite{ligo91}, Abramovici {\em et
al.\/}\cite{ligo92}).  LAGOS will be most sensitive to perturbed
black holes in the range $10^6\,{\rm M}_\odot\alt M\alt
10^7\,{\rm M}_\odot$, where $S_n\simeq10^{-42}\,{\rm Hz}^{-1}$
({\em cf.\/} Faller {\em et al.}\cite{lagos}). Consequently
\begin{equation}
\overline{\rho^2}^{1/2}\simeq
5.8 G(a)
\left\{\begin{array}{ll}
\left(\epsilon\over4\times10^{-8}\right)^{1/2}
\left(3\,{\rm Mpc}\over r\right)
\left(M\over50\,{\rm M}_\odot\right)^{3/2}
\left(10^{-48}{\rm Hz}^{-1}\over S_n\right)^{1/2}
&\quad\mbox{LIGO}\\
\left(\epsilon\over5\times10^{-9}\right)^{1/2}
\left(3\,{\rm Gpc}\over r\right)
\left(M\over10^6\,{\rm M}_\odot\right)^{3/2}
\left(10^{-42}{\rm Hz}^{-1}\over S_n\right)^{1/2}
&\quad\mbox{LAGOS}
\end{array}
\right.
\label{eqn:lagos-ligo}
\end{equation}
\narrowtext
where
\begin{equation}
G(a) \equiv {37\over200}\left[{17Q^2\over
F^2\left(1+4Q^2\right)}\right]^{1/2}. \label{defn:g(a)}
\end{equation}

For frequencies outside of the range $100\,{\rm Hz}$ to
$200\,{\rm Hz}$, the LIGO PSD $S_n$ scales with frequency: for
frequencies greater than approximately $200$~Hz (corresponding to
$M\alt50\,{\rm M}_\odot$), $S_n$ scales as $f^{2}$ ({\em cf.\/}
Thorne\cite{300yrs}, Krolak, Lobo
\& Meers\cite{klm}), and for frequencies less than 100~Hz
($M\agt100\,{\rm M}_\odot$) it scales as $f^{-4}$ ({\em cf.\/}
Dhurandhar, Krolak, \& Lobo\cite{dhurandhar}). Similarly, for
frequencies outside the range $10^{-3}\,{\rm Hz}$ to $10^{-2}\,{\rm Hz}$
the LAGOS PSD $S_n$ scales with frequency:
for $f\agt10^{-2}\,{\rm Hz}$ it scales as $f^2$,
and for $f\alt10^{-3}\,{\rm Hz}$ it scales as $f^{-4}$
({\em cf.\/} Faller {\em et
al.\/}\cite{lagos}). Consequently
\begin{equation}
\overline{\rho^2} \propto \left\{
\begin{array}{ll}
M^{-1}&\qquad M\agt100\,{\rm M}_\odot\\
M^{5}&\qquad M\alt50\,{\rm M}_\odot
\end{array}\right.
\end{equation}
for LIGO and
\begin{equation}
\overline{\rho^2} \propto \left\{
\begin{array}{ll}
M^{-1}&\qquad M\agt10^7\,{\rm M}_\odot\\
M^{5}&\qquad M\alt10^6\,{\rm M}_\odot
\end{array}\right.
\end{equation}
for LAGOS.

Little is known about the rate of, or the energy radiated during,
black hole formation ({\em cf.\/} Rees\cite{rees83}, Kochanek,
Shapiro \& Teukolsky\cite{kst}); however, owing to the extreme
sensitivity of both the LAGOS and LIGO detectors, it seems a
conservative estimate that the formation of a black hole of mass
$10^6\,{\rm M}_\odot\alt M\alt10^7\,{\rm M}_\odot$ anywhere in
the universe will be detectable by LAGOS, and the formation of
black holes with $50\,{\rm M}_\odot\alt M\,\alt100{\rm M}_\odot$
will be observable in LIGO at least to the distance of the Virgo
cluster ($\sim$10~Mpc).  Additionally, note that the energy
radiated in the $l=2$ mode during the radial infall of a test
body (mass $m$) onto a Schwarzschild black hole (mass $M$) is
given by
\begin{equation}
\Delta E = \epsilon M \simeq10^{-2} {m^2\over M},
\end{equation}
(Davis, Ruffini, Press, \& Price\cite{davis}, Oohara \&
Nakamura\cite{oohara}; similar results hold for Kerr black holes:
Sasaki
\& Nakamura\cite{sasaki}, Kojima \& Nakamura\cite{kojima}).
Consequently, the capture of a solar mass compact object ({\em
e.g.,\/} a black hole or neutron star) onto a black hole of mass
$10^6$--$10^7\,{\rm M}_\odot$ (corresponding to
$\epsilon\simeq10^{-14}\mbox{--}10^{-16}$) may also be observable
to a distance of $3\,{\rm Mpc}$ ({\em cf.\/}
eqn.~\ref{eqn:lagos-ligo})

Figure \ref{fig:snr} shows the factor $G(a)$ ({\em cf.\/}
eqn.~\ref{defn:g(a)}) as a function of $a$. This figure may also
be regarded as a plot of $\rho(a)$ for fixed $M$, $\epsilon$,
$r$, $S_n$, and detector-source orientation.  With this
interpretation, note how $\rho$ {\em decreases\/} with increasing
$a$. The reason for this behavior is that at fixed $M$, the
frequency $f$ and damping timescale $\tau$ both increase with
$a$; consequently, a signal of smaller amplitude ({\em i.e.,\/}
smaller $\rho^2$) will yield the same radiated energy.

\subsection{Precision of measurement}

While the parameters $Q$ and $f$ are convenient for
characterizing the detector response, it is the determination of
$a$ and $M$ that is of direct physical interest. If the perturbed
black hole is also observed electromagnetically ({\em e.g.,\/} if
it is the result of the gravitational collapse of a star in a
type II supernova), then determination of $V$ and $T$ may also be
interesting. Regardless, we are more interested in the covariance
matrix for the parameters $\{a,\,M,\,V,\,T\}$ than for the
parameters $\{Q,\,f,\,V,\,T\}$. It turns out, however, that it is
simpler to first determine the covariance matrix for the
parameterization $\{Q,\,f,\,V,\,T\}$.

To find the covariance matrix for the parameterization
$\{a,\,M,\,V,\,T\}$, first define the three dimensionless
parameters $\epsilon'$, $\xi'$ and $\zeta'$ by
\begin{mathletters}
\begin{eqnarray}
\widehat{f}\epsilon' &\equiv& \widetilde{f}-\widehat{f},\\
\widehat{V}\xi' &\equiv& \widetilde{V}-\widehat{V},\\
\zeta' &\equiv& \widehat{f}\left(\widetilde{T}-\widehat{T}\right),
\end{eqnarray}
\end{mathletters}
and evaluate ${\cal C}_{ij}'^{-1}$ for the parameterization
$\{Q,\,\epsilon',\,\xi',\,\zeta'\}$:
\FL
\begin{equation}
2\left(
{\arraycolsep=2\arraycolsep
\begin{array}{llll}
\left<{\partial h\over\partial Q},
{\partial h\over\partial Q}\right>&
\left<{\partial h\over\partial Q},
{\partial h\over\partial f}\right>f&
\left<{\partial h\over\partial Q},
{\partial h\over\partial V}\right>V&
\left<{\partial h\over\partial Q},
{\partial h\over\partial T}\right>
{1\over f}\\
&
\left<{\partial h\over\partial f},
{\partial h\over\partial f}\right>f^2&
\left<{\partial h\over\partial f},
{\partial h\over\partial V}\right>fV&
\left<{\partial h\over\partial f},
{\partial h\over\partial T}\right>\\
&
&
\left<{\partial h\over\partial V},
{\partial h\over\partial V}\right>V^2&
\left<{\partial h\over\partial V},
{\partial h\over\partial T}\right>
{V\over f}\\
&
&
&
\left<{\partial h\over\partial T},
{\partial h\over\partial T}\right>
{1\over f^2}
\end{array}}
\right).\label{eqn:cprime}
\end{equation}
The components of ${\cal C}_{ij}^{\prime-1}$ appearing in
equation
\ref{eqn:cprime} are
\begin{mathletters}
\begin{eqnarray}
\left<{\partial h\over\partial Q},
{\partial h\over\partial Q}\right>
&=& {3+6Q^2+8Q^4\over2Q^2\left(1+4Q^2\right)^2}\rho^2 \\
\left<{\partial h\over\partial Q},
{\partial h\over\partial f}\right>f
&=& -{3+4Q^2\over4Q\left(1+4Q^2\right)}\rho^2 \\
\left<{\partial h\over\partial Q},
{\partial h\over\partial V}\right>V
&=& -{3+4Q^2\over12Q\left(1+4Q^2\right)}\rho^2 \\
\left<{\partial h\over\partial Q},
{\partial h\over\partial T}\right>
{1\over f} &=& {\pi\rho^2\over4Q^2} \\
\left<{\partial h\over\partial f},
{\partial h\over\partial f}\right>f^2
&=& \left({1\over2}+Q^2\right)\rho^2\\
\left<{\partial h\over\partial f},
{\partial h\over\partial V}\right>fV
&=& {\rho^2\over12}\\
\left<{\partial h\over\partial f},
{\partial h\over\partial T}\right>
&=& -{\pi\rho^2\left(1+4Q^2\right)\over 4Q} \\
\left<{\partial h\over\partial V},
{\partial h\over\partial V}\right>V^2
&=&{\rho^2\over18}\\
\left<{\partial h\over\partial V},
{\partial h\over\partial T}\right>
{V\over f} &=&0\\
\left<{\partial h\over\partial T},
{\partial h\over\partial T}\right>
{1\over f^2} &=&{\pi^2\rho^2\left(1+4Q^2\right)\over 2Q^2}.
\end{eqnarray}
\end{mathletters}
The components of the covariance matrix ${\cal C}'_{ij}$ are
\begin{mathletters}
\begin{eqnarray}
{\cal C}'_{QQ} &=& {4Q^4+3A^2+1\over2Q^2\rho^2}\\
{\cal C}'_{Q\epsilon'} &=& {1\over2Q^{3}\rho^2}\\
{\cal C}'_{Q\xi'} &=& {3(4Q^4+5Q^{2}+1)\over2 Q^3\rho^2}\\
{\cal C}'_{Q\zeta'} &=& -{1\over2\pi\rho^2}\\
{\cal C}'_{\epsilon'\epsilon'}
&=& {1-2Q^2\left(1-4Q^2\right)\over
2Q^4\left(1+4Q^2\right)\rho^2}\\
{\cal C}'_{\epsilon'\xi'}
&=& {3\left(1-Q^2\right)\over2Q^4\rho^2}\\
{\cal C}'_{\epsilon'\zeta'} &=&
-{1\over2\pi\rho^2}{1-4Q^2\over Q\left(1+4Q^2\right)}\\
{\cal C}'_{\xi'\xi'}
&=& {9\left(1+2Q^2\right)^2\over2Q^4\rho^2}\\
{\cal C}'_{\xi'\zeta'}
&=& -{3\over2\pi Q\rho^2}\\
{\cal C}'_{\zeta'\zeta'}
&=& {2Q^2\over\pi^2\left(1+4Q^2\right)\rho^2}.
\end{eqnarray}
\end{mathletters}

Now define the three dimensionless parameters $\epsilon$, $\xi$,
and $\zeta$ by
\begin{mathletters}
\begin{eqnarray}
\widehat{M}\epsilon &\equiv& \widetilde{M}-\widehat{M}\\
\widehat{V}\xi &\equiv& \widetilde{V}-\widehat{V}\\
\widehat{M}\zeta &\equiv& \widetilde{T}-\widehat{T}.
\end{eqnarray}
\end{mathletters}
The covariance matrix ${\cal C}_{ij}$ for the parameterization
$\{a,\,\epsilon,\,\xi,\,\zeta\}$ is given in terms of ${\cal
C}'_{ij}$ by
\begin{equation}
{\cal C}_{ij} = \sum_{k,l}{\cal J}^{-1}_{ik}{\cal C}'_{kl}{\cal
J}^{-1}_{lj},
\end{equation}
where the symmetric matrix ${\cal J}_{ij}$ is given by
\begin{equation}
{\cal J} = \left(
{\arraycolsep=2\arraycolsep
\begin{array}{cccc}
{dQ\over da}&-{1\over fM}{df\over da}&0&0\\
&\left(fM\right)^2&0&0\\
&&1&0\\
&&&\left(fM\right)^2
\end{array}}
\right).
\end{equation}
Like ${\cal C}'_{ij}$, the matrix ${\cal C}_{ij}$ is a function
only of $\widehat{a}$ and $\rho^2$, and has the elements
\begin{mathletters}
\begin{eqnarray}
{\cal C}_{aa}
&=& {\left(1+2Q^2\right)\left(1+4Q^2\right)\over 2Q^2{Q'}^2}
{1\over\rho^2},
\label{eqn:sigma_a}\\
{\cal C}_{\epsilon\epsilon} &=&
\left\{
{
  \left[
    QF'\left(1+2Q^2\right)\left(1+4Q^2\right)
    -2FQ'
  \right]F'
  \over
  2 Q^3{Q'}^2F^2}
\right.\nonumber\\
&&
\left.\qquad{}+
{
  \left.
    1-2Q^2+8Q^4
  \right.\over2 Q^4\left(1+4Q^2\right)
}
\right\}
{1\over\rho^2},
\label{eqn:sigma_M}\\
{\cal C}_{\xi\xi}
&=& {9\left(1+2Q^2\right)^2\over 2 Q^4\rho^2},
\label{eqn:sigma_V}\\
{\cal C}_{\zeta\zeta}
&=& {8Q^2\over\left(1+4Q^2\right)F^2\rho^2},
\label{eqn:sigma_T}\\
{\cal C}_{a\epsilon} &=&
{
  Q\left(1+2Q^2\right)\left(1+4Q^2\right)F' - FQ'
  \over 2 FQ^3 {Q'}^2\rho^2
},
\label{eqn:corr_aM}\\
{\cal C}_{a\xi} &=&
{3\left(1+4Q^2\right)\left(1+Q^2\right)\over2Q^3{Q'}^2\rho^2},
\label{eqn:corr_aV}\\
{\cal C}_{a\zeta} &=& -\left( FQ'\rho^2\right)^{-1}
\label{eqn:corr_aT}\\
{\cal C}_{\epsilon\xi} &=& {
3\left[
  Q\left(1+Q^2\right)\left(1+4Q^2\right)F'
  + \left(Q^2-1\right)FQ'
\right]\over2Q^4FQ'\rho^2
},\nonumber\\
\label{eqn:corr_MV}\\
{\cal C}_{\epsilon\zeta} &=& {
\left(1-4Q^2\right)FQ'-Q\left(1+4Q^2\right)F'
\over Q\left(1+4Q^2\right)F^2Q'\rho^2},
\label{eqn:corr_MT}\\
{\cal C}_{\xi\zeta} &=& -{3\over QF\rho^2},
\label{eqn:corr_VT}
\end{eqnarray}
\end{mathletters}
where $F(a)$ and $Q(a)$ are given by equations \ref{eqn:f(a)} and
\ref{eqn:q(a)}. Finally, in terms of these coefficients, we have
({\em cf.\/} eqns.~\ref{defn:sigma} and \ref{defn:correlation})
\begin{mathletters}
\begin{eqnarray}
\sigma^2_M &=&
\widehat{M}^2\sigma^2_\epsilon\label{eqn:fig2-first}\\
\sigma^2_V &=& \widehat{V}^2\sigma^2_\xi\\
\sigma^2_T &=& \widehat{M}^2\sigma^2_\zeta\label{eqn:fig2-last}\\
r_{aM} &=& r_{a\epsilon}\label{eqn:fig3a}\\
r_{aV} &=& r_{a\xi}\label{eqn:fig3b-first}\\
r_{aT} &=& r_{a\zeta}\label{eqn:fig3b-last}\\
r_{MV} &=& r_{\epsilon\xi}\label{eqn:fig3c-first}\\
r_{MT} &=& r_{\epsilon\zeta}\\
r_{VT} &=& r_{\xi\zeta}.\label{eqn:fig3c-last}
\end{eqnarray}
\end{mathletters}

The results for the standard deviations $\sigma_a$ and
$\sigma_M/M$ and correlation coefficient $r_{aM}$ found
semi-numerically in Echeverria\cite{echeverria} (his
eqns.~4.10a--c and table~II) are approximations to the analytic
results found here in equations~\ref{eqn:sigma_a},
\ref{eqn:fig2-first} and \ref{eqn:fig3a}. Additionally, we give
analytic forms of the other variances and correlations
coefficients.

Figure~\ref{fig:sigma-const-rho} shows $\sigma_a$,
$\sigma_M/\widehat{M}$, $\sigma_V/\widehat{V}$, and
$\sigma_T/\widehat{M}$ (eqns.~\ref{defn:sigma},
\ref{eqn:sigma_a}--\ref{eqn:sigma_T}), normalized by $\rho$ as
shown, as functions of the measured $a$.  Note how for
$a\alt0.8$, the angular momentum parameter is determined less
precisely than the mass. Figures~\ref{fig:r}a--c shows the six
correlation coefficients ({\em cf.\/}
eqns.~\ref{defn:correlation},
\ref{eqn:corr_aM}--\ref{eqn:corr_VT}) as functions of the
measured $a$.  These are independent of $\rho$.  In
figures~\ref{fig:r}a--c note how $\delta a$ and $\delta M$ are
highly correlated so and are not statistically independent
parameters: for a complete discussion of this point, see
Echeverria\cite{echeverria} \S{IV}.

Figure \ref{fig:sigma-const-rho} shows the standard deviations
for fixed $\rho$. It is also useful to consider these same
quantities for fixed $M$, $r$, $\epsilon$ and $S_n$ as was done
in equation \ref{eqn:rho-epsilon}, \ref{eqn:lagos-ligo} and
figure~\ref{fig:snr}.  Defining $\rho_0$ by
\begin{equation}
\rho = \rho_0 G(a),\label{eqn:rho0}
\end{equation}
where $G(a)$ is given in equation \ref{defn:g(a)},
figure~\ref{fig:sigma-const-epsilon} shows $\sigma_a$,
$\sigma_M/M$, $\sigma_V/V$ and $\sigma_T/M$ normalized by
$\rho_0$ (for LAGOS and LIGO, $\rho_0$ is given by equation
\ref{eqn:lagos-ligo}) and as functions of $a$. Note the
difference between figures~\ref{fig:sigma-const-rho} and
\ref{fig:sigma-const-epsilon}: in the first case, the SNR is held
constant while in the second the the energy of the perturbation
is held constant. In the second case, the precision with which
$a$ and $M$ can be determined does not increase as rapidly with
$a$ as in the first case, the precision with which $T$ can be
determined is independent of $a$, and the precision with which
the amplitude can be determined {\em decreases\/} with increasing
$a$.

The elements of ${\cal C}^{-1}_{ij}$ fully determine the
distribution $p[m(\mbox{\boldmath$\mu$})|g]$ and the volumes
$V(P)$.  Generally we will have no interest in $V$ and $T$, in
which case we integrate the distribution over all $T$ and $V$ to
find the two-dimensional distribution
\begin{eqnarray}
\lefteqn{p[m(a,M)|g]=}\qquad\nonumber\\
&&{\exp\left[
-{1\over2\left(1-r_{aM}^2\right)}
\left(
{\Delta M\over\sigma_M^2}+
{\Delta a\over\sigma_a^2}
-2{\Delta a\Delta Mr_{aM}\over\sigma_a\sigma_M}
\right)
\right]\over
2\pi\sigma_a\sigma_M\left(1-r^2_{aM}\right)^{1/2}
}\nonumber\\
\end{eqnarray}
where
\begin{mathletters}
\begin{eqnarray}
\Delta M &\equiv& M-\widehat{M}\\
\Delta a &\equiv& a-\widehat{a}.
\end{eqnarray}
\end{mathletters}

\section{Discussion}
\label{sec:discussion}

The earlier results of Echeverria\cite{echeverria} on the
precision of measurement are restricted to the case of large
$\rho$ where the distribution of $\delta\mbox{\boldmath$\mu$}$ is
well-approximated by a Gaussian, though there is no discussion of
what constitutes a sufficiently large SNR. Additionally, those
results do not provide any guidance for estimating the precision
with which the amplitude of the signal can be measured.  Finally,
there is no clear connection drawn between the measurement of the
$\widehat{\mbox{\boldmath$\mu$}}$, the estimates of
$\overline{\delta\mu_i\delta\mu_j}$, and the probability that
$|\widetilde{\mbox{\boldmath$\mu$}}
-\widehat{\mbox{\boldmath$\mu$}}|^2\le\overline{\delta\mu^2_i}$.
The restriction to strong signals is required because of the
expansion of Echeverria's\cite{echeverria} expression for $\rho$
in a power-series about the ``measured'' parameters and also
because the methods described fail to take into account prior
knowledge about the distribution of the parameters [{\em i.e.,\/}
the $p(\mbox{\boldmath$\mu$})$].  This prior knowledge plays an
important role when the distribution of the
$\delta\mbox{\boldmath$\mu$}$ is not uniform or the SNR is small.

On the other hand, the maximum likelihood analysis described in
\S\ref{sec:detection-and-measurement} is applicable for all SNR
(though the approximate techniques discussed at the end of
\S\ref{sec:sensitivity} are appropriate only when the
distribution of the $\delta\mbox{\boldmath$\mu$}$ is
well-approximated by a Gaussian).  It does not elevate any
parameter to a special status: the amplitude of the signal and
its precision are determined in the same way that all other
signal parameters and their precision are determined.  Finally,
it gives clear meaning to the measured parameters
$\widehat{\mbox{\boldmath$\mu$}}$ and the precision of
measurement by providing the probability distribution of the
$\delta\mbox{\boldmath$\mu$}$.

A complete discussion of how the optimal filter techniques of
Echeverria\cite{echeverria} are related to the maximum likelihood
analysis presented here can be found in Echeverria \& Finn
\cite{ech-finn}.

\section{Conclusions}
\label{sec:conclusions}

In the analysis of the results of an observation made in a
detector, it is useful to distinguish between {\em detection\/}
and {\em measurement.} The analysis involved in detection refers
only to the presence or absence of a signal characterizing a
particular {\em class\/} of sources to which the detector is
sensitive ({\em e.g.,\/} perturbed black holes). A particular
source of this class is described by a set of parameters: {\em
e.g.,\/} among the parameters describing the signal from a
perturbed black hole is the black hole mass and angular momentum.
Detection addresses only whether a signal of this class is
present in the observed output of the detector, and not the
particular values of the parameters that best describe the
signal.

Measurement follows detection: it refers to the determination of
the values of the parameters that best characterize the
particular signal {\em assumed to be present in the detector
output\/} (it only makes sense to speak of measuring the
parameters of a real signal). For example, once we have concluded
that we have {\em detected\/} the signal from the formation of a
black hole, then we can go on to {\em measure\/} the black hole
mass and angular momentum.

In order to determine whether the observed output of a detector
includes a signal from a given class of sources, we saw how to
calculate the probability that the detector output is consistent
with the presence of the signal.  That probability depends on the
characteristics of the detector noise, the observed detector
output, and a parameterized model of the detector response to the
signal. In addition, it depends on several {\em a~priori\/}
probabilities that must be assessed.  When the calculated
probability exceeds a certain threshold then we say that the we
have detected a signal. Setting the threshold requires careful
consideration of the relative severity of falsely claiming a
detection and incorrectly rejecting a signal.

To determine the values of the parameters that characterize the
detected signal, we saw how to construct the probability
distribution that describes how likely different
parameterizations {\boldmath$\mu$} are.  We identified
$\widehat{\mbox{\boldmath$\mu$}}$ as the mode of the
distribution, {\em i.e.,\/} the parameterization that maximized
the probability density, or the {\em most likely\/}
parameterization.  Owing to detector noise,
$\widehat{\mbox{\boldmath$\mu$}}$ differs in a random fashion
from the unknown $\widetilde{\mbox{\boldmath$\mu$}}$ that
actually describes the signal. We characterized our uncertainty
over the actual description of the signal by specifying a volume
$V(P)$ in the parameter phase space, centered on
$\widehat{\mbox{\boldmath$\mu$}}$, such that the
$\widetilde{\mbox{\boldmath$\mu$}}\in V(P)$ with probability $P$.

We then proceeded to exploit these techniques to {\em
anticipate\/} the precision with which the parameterization of a
particular signal can be determined by a given detector: {\em
i.e.,\/} we evaluated the {\em sensitivity\/} of the detector to
the signal from a class of sources.

To do so, we found the probability distribution of
$\widetilde{\mbox{\boldmath$\mu$}} -
\widehat{\mbox{\boldmath$\mu$}}$ and defined volumes $V(P)$ in
phase space such that $\widetilde{\mbox{\boldmath$\mu$}}\in V(P)$
with probability $P$. These volumes determine the precision with
which we {\em expect\/} we can determine the signal parameters in
a real observation.  In the interesting limit of a strong signal
the anticipated probability distribution of
$\widetilde{\mbox{\boldmath$\mu$}} -
\widehat{\mbox{\boldmath$\mu$}}$ for fixed
$\widehat{\mbox{\boldmath$\mu$}}$ is close to Gaussian and the
associated volumes $V(P)$ are ellipsoids. In this limit we found
approximate techniques for determining the size and orientation
of this ellipsoid. Both the exact and approximate expressions
developed provide a powerful means of studying the sensitivity of
a proposed detector or detector configuration to a source of
gravitational radiation. These techniques are currently being
employed to study the sensitivity of the LIGO detectors to binary
coalescence\cite{finn-chernoff}, precessing axisymmetric neutron
stars\cite{finn92b}, and non-axisymmetric neutron
stars\cite{finn92c}.

As an example of the process of measurement, we evaluated the
variance in the mass and angular momentum of a perturbed black
hole as determined by observations in a gravitational wave
detector. These results improve upon those found earlier ({\em
cf.\/} Echeverria \cite{echeverria}), and we discussed the origin
of the differences.

The LIGO detector, currently under construction, and the LAGOS
detector, currently being designed, are both very sensitive to
gravitational radiation from perturbed black holes. A
perturbation of a $50\,{\rm M}_\odot$--$100\,{\rm M}_\odot$ black
hole that radiates as little as $10^{-7}$ of the black hole mass
should be observable with LIGO at the distance of the Virgo
cluster of galaxies, and a perturbation of a $10^6\,{\rm
M}_\odot$--$10^7\,{\rm M}_\odot$ black hole that radiates as
little as $10^{-8}$ of the black hole mass should be observable
by LAGOS throughout the Universe ({\em cf.\/}
eqn.~\ref{eqn:lagos-ligo}).

\acknowledgments

I am glad to thank David Chernoff, Fernando Echeverria, Eanna
Flannagan, and Kip Thorne for helpful conversations. I also
gratefully acknowledge the support of the Alfred P.\ Sloan
Foundation. This work was supported by a grant from the National
Aeronautics and Space Administration (NASA grant NAGW-2936).

\mediumtext
\figure{The expected signal-to-noise ratio (SNR) of the $l=m=2$
mode of a perturbed black hole as a function of the angular
momentum parameter $a$. The dependence of the SNR on the black
hole mass, distance, total energy radiated, and the detector
noise PSD has been scaled out, leaving only the dependence on the
angular momentum parameter. For more details see equation
\ref{eqn:lagos-ligo} and surrounding text.\label{fig:snr}}

\figure{The expected standard deviation of the black hole angular
momentum parameter ($\sigma_a$), mass ($\sigma_M/M$), initial
moment of perturbation ($\sigma_T/M$), and perturbation amplitude
($\sigma_V/V$) as a function of the angular momentum parameter
$a$. The dependence of these standard deviations on the
signal-to-noise ratio (SNR) $\rho$ has been scaled out as shown.
For more details see equations \ref{eqn:sigma_a},
\ref{eqn:fig2-first}--\ref{eqn:fig2-last} and surrounding
text.\label{fig:sigma-const-rho}}

\figure{The correlation coefficients for errors in the angular
momentum parameter $a$, mass $M$, initial moment of perturbation
$T$, and perturbation amplitude $V$ as a function of angular
momentum parameter.  For more details see equations
\ref{eqn:fig3a}--\ref{eqn:fig3c-last} and surrounding
text.\label{fig:r}}

\figure{Like figure 2, except that the total energy radiated by
the perturbation is held fixed instead of the SNR. Compare with
figures \ref{fig:snr} and \ref{fig:sigma-const-rho}. For more
details see \S\ref{sec:bhringdown}.
\label{fig:sigma-const-epsilon}}


\begin{references}
%
\bibitem[*]{author}Alfred P.\ Sloan Research Fellow.
%
\bibitem{ligo91}R.\ E.\ Vogt, {\em The U.\ S. LIGO Project,\/}
preprint LIGO 91-7 (1991).
%
\bibitem{ligo92}
A.\ Abramovici, W.\ E.\ Althouse, R.\ W.\ P.\ Drever, Y.\ G\"ursel,
S.\ Kawamura, F.\ J.\ Raab, D.\ Shoemaker, L.\ Sievers,
R.\ E.\ Spero, K.\ S.\ Thorne, R.\ E.\ Vogt, R.\ Weiss, S.\ E.\ Whitcomb,
\& M.\ E.\ Zucker, Science {\bf 256,} 325--333.
%
\bibitem{virgo}
C. Bradaschia {\em et al.,\/} Nucl. Instrum. \& Methods {\bf
A289,} 518 (1990).
%
\bibitem{lagos}
J.\ E.\ Faller, P.\ L.\ Bender, J.\ L.\ Hall, D.\ Hils, R.\ R.\ Stebbins,
\& M.\ A.\ Vincent, Adv.\ Space Res.\ (COSPAR) {\bf 9,}
(9)107--(9)11 (1989).
%
\bibitem{echeverria}
F.\ Echeverria, Phys.\ Rev.\ {\bf D40,} 3194 (1989).
%
\bibitem{ech-finn}
F.\ Echeverria \& L.\ S.\ Finn, in preparation (1992).
%
\bibitem{wainstein}
L.\ A.\ Wainstein, \& L.\ D.\ Zubakov, {\em Extraction of Signals
from Noise\/} (Prentice Hall, Englewood Cliffs, New Jersey 1962).
%
\bibitem{winkler}
R.\ L.\ Winkler, {\em An Introduction to Bayesian Inference and
Decision\/} (Holt, Rinehart and Winston Inc., New York 1972).
%
\bibitem{mathews}
J.\ Mathews \& R.\ L.\ Walker, {\em Mathematical Methods of
Physics\/} (Benjamin/Cummings, Menlo Park, California 1970).
%
\bibitem{kittel}
C.\ Kittel, {\em Elementary Statistical Physics\/} (John Wiley \&
Sons, New York 1958).
%
\bibitem{golub}
G.\ H.\ Golub \& C.\ F.\ Van Loan, {\em Matrix Computations, 2nd
Edition\/} (Johns Hopkins University Press, Baltimore 1989).
%
\bibitem{vinet}
J.-Y.\ Vinet, B.\ Meers, C.\ N.\ Man, \& A.\ Brillet,  Phys.\
Rev.\ {\bf D38,} 433 (1988).
%
\bibitem{meers}
B.\ J.\ Meers, Phys.\ Rev.\ {\bf D38,} 2317 (1988).
%
\bibitem{klm}
A.\ Krolak, J.\ A.\ Lobo, \& B.\ Meers, Phys.\ Rev.\ {\bf D43,} 2470
(1991).
%
\bibitem{300yrs}
K.\ S.\ Thorne, in {\em 300 Years of Gravitation,\/} edited by S.\
W.\ Hawking \& W.\ Israel (Cambridge University Press, Cambridge
1987) pgs.~330--458.
%
\bibitem{dhurandhar}
S.\ V.\ Dhurandhar, A.\ Krolak, \& J.\ A.\ Lobo, Mon.\ Not.\ R.\
astr.\ Soc.\ {\bf 237,\/} 333-340 (1989).
%
\bibitem{rees83}
M.\ J.\ Rees, in {\em Gravitational Radiation,} edited by N.\
Deruelle \& T.\ Piran (North-Holland Publishing Company,
Amsterdam 1983) pgs.~297--320.
%
\bibitem{kst}
C.\ S.\ Kochanek, S.\ L.\  Shapiro, \& S.\ A.\ Teukolsky, Ap.\ J.\
{\bf 320,} 73--84 (1987).
%
\bibitem{davis}
M.\ Davis, R.\  Ruffini, W.\ H.\ Press, \& R.\ H.\ Price,  Phys.\
Rev.\ Lett.\ {\bf 27,} 1466--1469 (1971).
%
\bibitem{oohara}
K.\ Oohara \& T.\ Nakamura, Prog.\ Theor.\ Phys.\ {\bf 70,}
757--771 (1983).
%
\bibitem{sasaki}
M.\ Sasaki \& T.\ Nakamura,  Prog.\ Theor.\ Phys.\ {\bf 67,}
1788--1809 (1982).
%
\bibitem{kojima}
Y.\ Kojima, \& T.\ Nakamura, Prog.\ Theor.\ Phys.\ {\bf 71,}
79--90 (1984).
%
\bibitem{finn-chernoff}
L.\ S.\ Finn \& D.\ Chernoff, in preparation (unpublished).
%
\bibitem{finn92b}
L.\ S.\ Finn, in preparation (unpublished).
%
\bibitem{finn92c}
L.\ S.\ Finn, in preparation (unpublished).
%
\end{references}
\end{document}